\documentclass[iop]{emulateapj}

\usepackage{graphicx}
\usepackage{txfonts}
\usepackage{booktabs}

\usepackage{natbib}
\shortauthors{Kishimoto et al.}
\shorttitle{Evidence for receding dust sublimation region in AGN}

\begin{document}

\title{Evidence for a receding dust sublimation region around a supermassive black hole}



\author{Makoto Kishimoto\altaffilmark{1},
       Sebastian F. H\"onig\altaffilmark{2,3},
       Robert Antonucci\altaffilmark{2}, 
       Rafael Millan-Gabet\altaffilmark{4},
       Richard Barvainis\altaffilmark{5},\\
       Florentin Millour\altaffilmark{6},
       Takayuki Kotani\altaffilmark{7}, 
       Konrad R.~W.~Tristram\altaffilmark{1},
       Gerd Weigelt\altaffilmark{1}
       }

\altaffiltext{1}{Max-Planck-Institut f\"ur Radioastronomie, Auf dem H\"ugel 69,
  53121 Bonn, Germany; \email{mk@mpifr-bonn.mpg.de}}
\altaffiltext{2}{Physics Department, University of California, Santa Barbara, CA 93106, USA}
\altaffiltext{3}{Institut f\"ur Theoretische Physik und Astrophysik, Christian-Albrechts-Universit\"at zu Kiel, Leibnizstr. 15, 24118, Kiel, Germany}
\altaffiltext{4}{Jet Propulsion Laboratory, California Institute of Technology, USA}
\altaffiltext{5}{National Science Foundation, 4301 Wilson Boulevard, Arlington, VA 22230, USA}
\altaffiltext{6}{Observatoire de la C\^ote d Azur, Departement FIZEAU, Boulevard de l'Observatoire, BP 4229, 06304, Nice Cedex 4, France}
\altaffiltext{7}{National Astronomical Observatory of Japan, 2-21-1 Osawa, Mitaka, Tokyo 181-8588, Japan}

%




\begin{abstract}

The near-IR emission in Type 1 AGNs is thought to be dominated by the thermal radiation from dust grains that are heated by the central engine in the UV/optical and are almost at the sublimation temperature. A brightening of the central source can thus further sublimate the innermost dust, leading to an increase in the radius of the near-IR emitting region. Such changes in radius have been indirectly probed by the measurements of the changes in the time lag between the near-IR and UV/optical light variation. Here we report direct evidence for such a receding sublimation region through the near-IR interferometry of the brightest Type 1 AGN in NGC4151. The increase in radius follows a significant brightening of the central engine with a delay of at least a few years, which is thus the implied destruction timescale of the innermost dust distribution. Compiling historic flux variations and radius measurements, we also infer the reformation timescale for the inner dust distribution to be several years in this galactic nucleus. More specifically and quantitatively, we find that the radius at a given time seems to be correlated with a long-term average of the flux over the previous several ($\sim$6) years, instead of the instantaneous flux. 
Finally, we also report measurements of three more Type 1 AGNs newly observed with the Keck interferometer, as well as the second epoch measurements for three other AGNs.

\end{abstract}


\maketitle


%

\section{Introduction}

The near-IR emission in the $K$-band (2.2 $\mu$m) in Type 1 AGNs is thought to be the thermal radiation emitted predominantly from the innermost dust grains, which are heated by UV/optical emission from the central engine or the putative accretion disk (AD). These grains are considered to be near the sublimation temperature. The size of this emission region was first probed by the reverberation measurements of the time lag of the variation in the near-IR from that in the UV/optical. The near-IR reverberation radii over a {\it sample} of objects have been shown to scale with the central engine's luminosity $L$ as $\propto L^{1/2}$ (\citealt{Oknyanskij01,Suganuma06}), which suggests that the innermost dust grain properties are indeed similar in different objects \citep{Barvainis87}. The reverberation radius in the brightest Seyfert 1 galaxy NGC4151 has also been reported to change as the central luminosity varies  \citep{Oknyanskij99,Oknyanskij06,Koshida09}, but not simply with $L^{1/2}$. 

The sublimation region has been directly resolved by near-IR interferometers, namely the Keck interferometer (KI) and VLTI/AMBER
\citep{Swain03,Kishimoto09KI,Pott10,Kishimoto11,Weigelt12}. In these observations, NGC4151 did not show any significant size variation, despite a significant change in the optical flux \citep{Pott10,Kishimoto11}. Here we report further near-IR interferometric measurements with the KI for NGC4151 in multiple epochs as well as those for six other targets. 
Together with our most recent VLTI/AMBER observations \citep{Kishimoto13}, we aim to investigate the size variation and the destruction/reformation timescales of the innermost dust distribution.

\begin{table*}
{\scriptsize

\caption[]{Summary of KI observations.}
\begin{tabular}{lccccccccccccccccccccc}
\hline
\hline
target & date & $B_{p}$ & PA        & $V^2$               & \multicolumn{2}{c}{$R_{\rm ring}^a$}    & $f_{\rm AD}^b$ & \multicolumn{2}{c}{$R_{\rm ring}$ corr.$^c$} & \multicolumn{5}{c}{flux (mJy)$^d$} \\
\cmidrule(rl){6-7} \cmidrule(rl){9-10} \cmidrule(rl){11-15}
       & (UT) & (m)     & ($\degr$) &                             & (mas)         & (pc)            &      & (mas)         & (pc) & Z & Y & J & H & K\\
\hline
\multicolumn{5}{l}{\it re-observed with KI}\\
NGC4151     & 2011-06-14 & 84.9 &  5.6 & 0.961$\pm$0.104                 & $<$ 0.60      & $<$ 0.052       & 0.14$\pm$0.05$^e$ & $<$ 0.66      & $<$ 0.057       \\
NGC4151     & 2012-05-03 & 84.9 & 22.9 & 0.714$\pm$0.079                 & 0.65$\pm0.09$ & 0.056$\pm$0.008 & 0.18$\pm$0.06     & 0.71$\pm$0.10 & 0.061$\pm$0.009 & 32.1 & 36.0 & 49.0 & 77.2 & 133. \\
NGC4151     & 2012-05-04 & 81.3 & 40.6 & 0.791$\pm$0.105                 &\multicolumn{5}{l}{(combined for fit above)} \\
3C273       & 2011-06-14 & 70.9 & 23.3 & 0.964$\pm$0.032                 & 0.27$\pm$0.14 & 0.75$\pm$0.39   & 0.28$\pm$0.09$^e$ & 0.32$\pm$0.17 & 0.88$\pm$0.47 \\
MRK231      & 2012-05-03 & 79.6 & 27.9 & 0.882$\pm$0.071                 & 0.45$\pm$0.15 & 0.38$\pm$0.13   & 0.15$\pm$0.05$^f$ & 0.47$\pm$0.17 & 0.39$\pm$0.14 \\
AKN120      & 2011-10-11 & 79.4 & 37.0 & 0.804$\pm$0.035                 & 0.59$\pm$0.06 & 0.38$\pm$0.04   & 0.17$\pm$0.05$^e$ & 0.63$\pm$0.07 & 0.41$\pm$0.04 \\
\hline
\multicolumn{5}{l}{\it new KI targets}\\
MRK509      & 2010-10-24 & 74.8 & 40.4 & 0.908$\pm$0.038                 & 0.42$\pm$0.09 & 0.28$\pm$0.06   & 0.17$\pm$0.06     & 0.45$\pm$0.11 & 0.30$\pm$0.07 \\
MRK509      & 2011-10-11 & 79.5 & 42.2 & 0.895$\pm$0.045                 & 0.43$\pm$0.10 & 0.29$\pm$0.07   & 0.17$\pm$0.06     & 0.46$\pm$0.11 & 0.31$\pm$0.07   & 11.7 & 12.6 & 16.2 & 28.0 & 52.9\\
MRK509      & 2011-10-11 & 59.0 & 26.0 & 0.923$\pm$0.056                 &\multicolumn{5}{l}{(combined for fit above)} \\
NGC7469$^g$ & 2010-10-24 & 84.9 & 40.0 & 1.064$\pm$0.066                 & $<$ 0.30      & $<$ 0.09        & 0.14$\pm$0.05     & $<$ 0.32      & $<$ 0.10        & 12.5 & 11.7 & 16.3 & 39.0 & 51.8\\
NGC7603$^g$ & 2011-10-11 & 75.2 & 32.1 & 0.965$\pm$0.048                 & $<$ 0.51      & $<$ 0.29        & 0.25$\pm$0.07     & $<$ 0.58      & $<$ 0.33        &  --  &  --  & 20.4 & 32.5 & 46.1\\
NGC7603$^g$ & 2011-10-11 & 68.9 & 22.3 & 0.934$\pm$0.060                 &\multicolumn{5}{l}{(combined for fit above)} \\
\hline
\multicolumn{5}{l}{\it fit results for combined data}\\
NGC4151 & \multicolumn{4}{c}{2011 KI data with VLTI 2011-05-15 data} & $<$ 0.63      & $<$ 0.054       & 0.14$\pm$0.05$^e$ & $<$ 0.68      & $<$ 0.058       \\
NGC4151 & \multicolumn{4}{c}{2012 KI data with VLTI 2012-05-10 data} & 0.68$\pm$0.08 & 0.058$\pm$0.006 & 0.18$\pm$0.06     & 0.75$\pm$0.09 & 0.064$\pm$0.008 \\
\hline
\end{tabular}
\\
$^a$ Thin-ring fit radius, incorporating conservative systematic uncertainty of 0.03 in $V^2$ (Sect.3.2 in \citealt{Kishimoto11}). Upper limits are at 95\% confidence, or 2-$\sigma$. 
$^b$ AD flux fraction at $K$-band. 
$^c$ After AD contribution correction.
$^d$ Point-source flux from 2D fit of Calar Alto images taken on 2012-07-04 (except for NGC7603 where 2MASS images taken on 2000-08-25 were used). Uncertainty is $\sim$5 \% (see sect.3 in \citealt{Kishimoto09KI}).
$^e$ \cite{Kishimoto11}. $^f$ \cite{Kishimoto09KI}. $^g$ The number of data blocks is only 5-10 (see Fig.\ref{fig_rawvis}), and thus less reliable than the others.
}
\label{tab_obs}
\end{table*}

\begin{figure*}
\centering
\includegraphics[width=16.0cm]{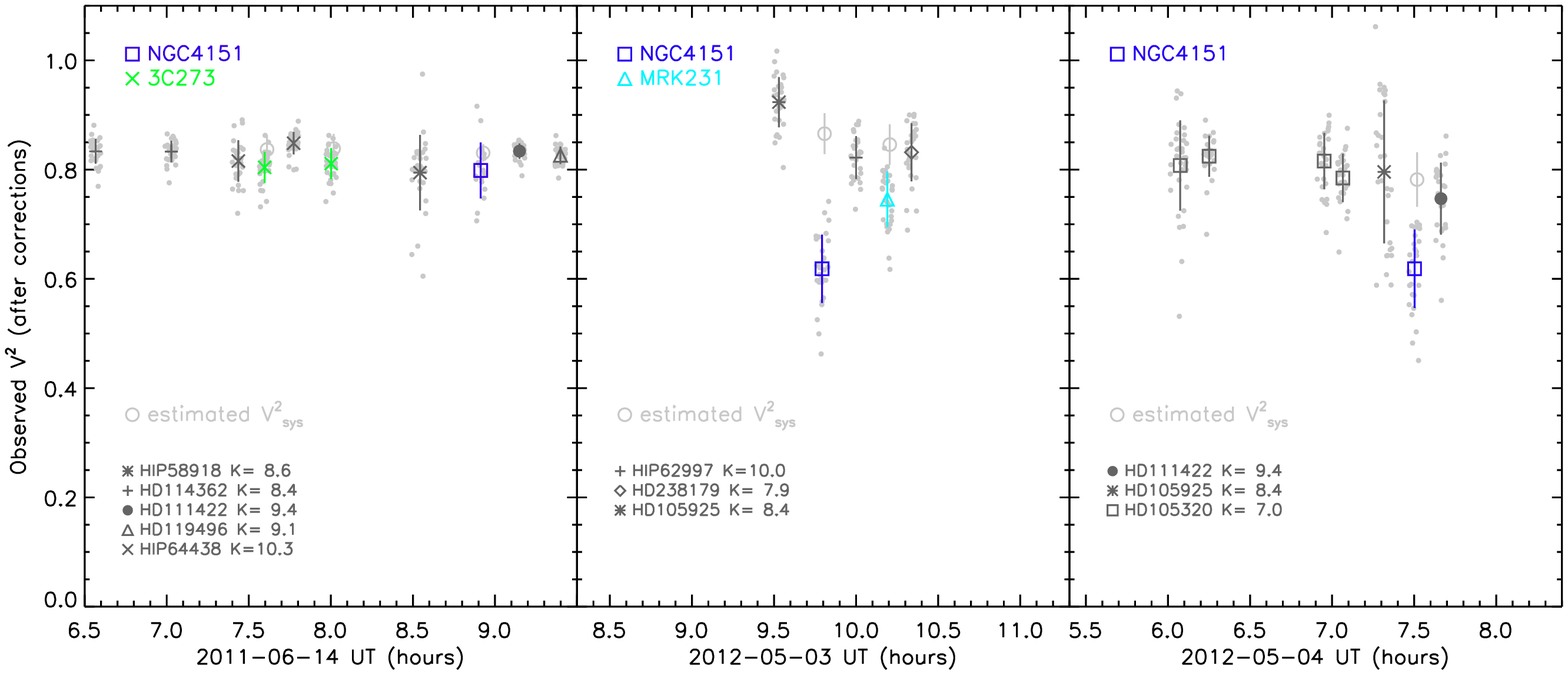}
\includegraphics[width=16.0cm]{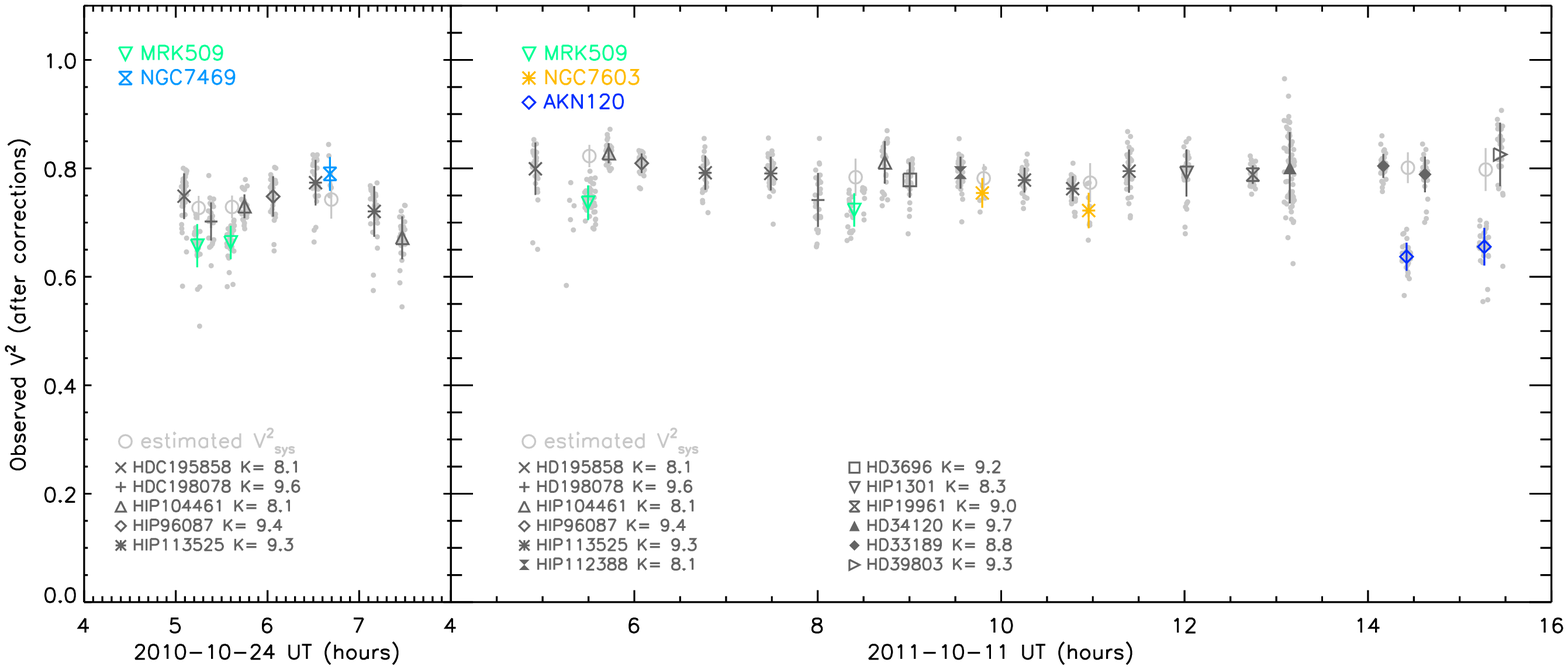}
\caption{Observed squared-visibility (after the correction for flux bias and flux ratio) obtained with KI.
Gray dots are individual measurements for blocks of 5 sec each. These blocks are averaged into scans, which are indicated by various symbols with error bars.}
\label{fig_rawvis}
\end{figure*}


\section{Observations and data reduction}\label{sec_obs}


We observed seven Type 1 AGNs in Oct 2010, Jun/Oct 2011, and May 2012 as listed in Table~\ref{tab_obs}, where three were new KI targets.
The data were reduced in the standard manner as described in \cite{Kishimoto09KI,Kishimoto11} using the standard software {\sf Kvis} and {\sf WbCalib}. Figure~\ref{fig_rawvis} shows the observed visibilities (after correction for flux bias and flux ratio\footnotemark[1]) of calibrators and targets in each night.
The data for NGC4151 in June 2011 were taken under unstable conditions and also quite close to the telescope pointing limit. The default extraction of the flux ratio measurements included the last measurement with one of the two beams showing no flux. Therefore, we excluded the last set of the flux measurements and recalculated the ratio correction manually. Here we conservatively incorporated the fluctuation of the ratio correction measurements into the final error of the calibrated visibility. 
All the calibrated visibilities 
are summarized in Table~\ref{tab_obs}.

\footnotetext[1]{http://nexsci.caltech.edu/software/KISupport/dataMemos}


To determine contemporaneous spectral energy distributions (SED), we obtained near-IR photometry under Director's Discretionary Time at Calar Alto Observatory for NGC4151, MRK509 and NGC7469. We used the instrument OMEGA2000, whose wide field of view provided simultaneous point-spread-function (PSF) measurements, enabling an accurate two-dimensional decomposition of the PSF and host galaxy component (Sect.3 of \citealt{Kishimoto09KI}). The results are summarized in Table~\ref{tab_obs}. Here we also include fluxes for NGC7603 obtained with the same procedure from 2MASS images.





\begin{figure}
\includegraphics[width=9cm]{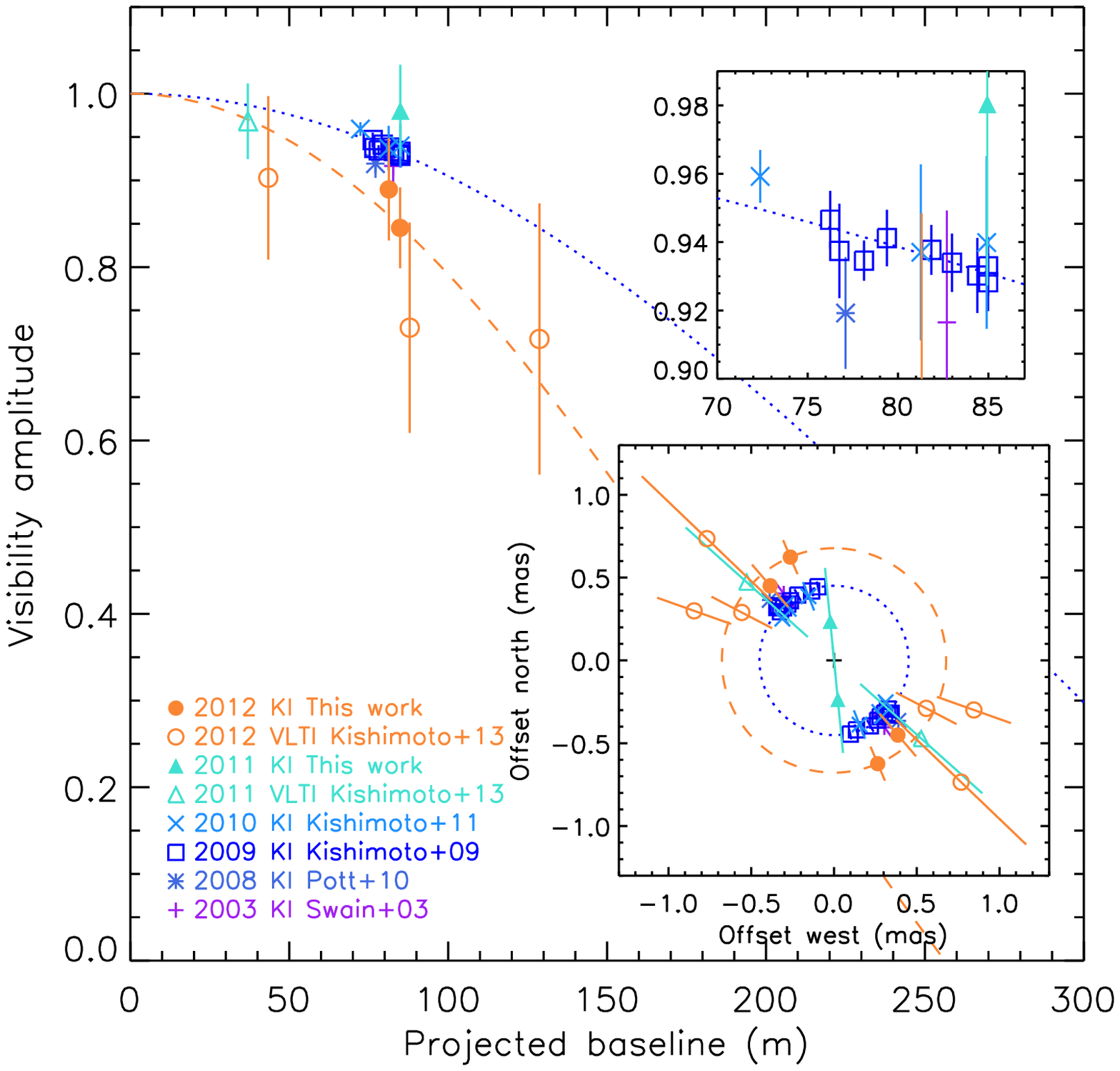}
\caption{$K$-band visibilities of NGC4151 with the KI and VLTI/AMBER. The top-inset enlarges the crowded data points. The bottom-inset shows the thin-ring radii, each plotted along the baseline PA. The best-fit ring models (without the AD component) are shown in dashed and dotted lines for 2012 and 2009 data, respectively.}
\label{fig_vis_ngc4151}
\end{figure}

\begin{figure*}
\centering
\includegraphics[width=\textwidth]{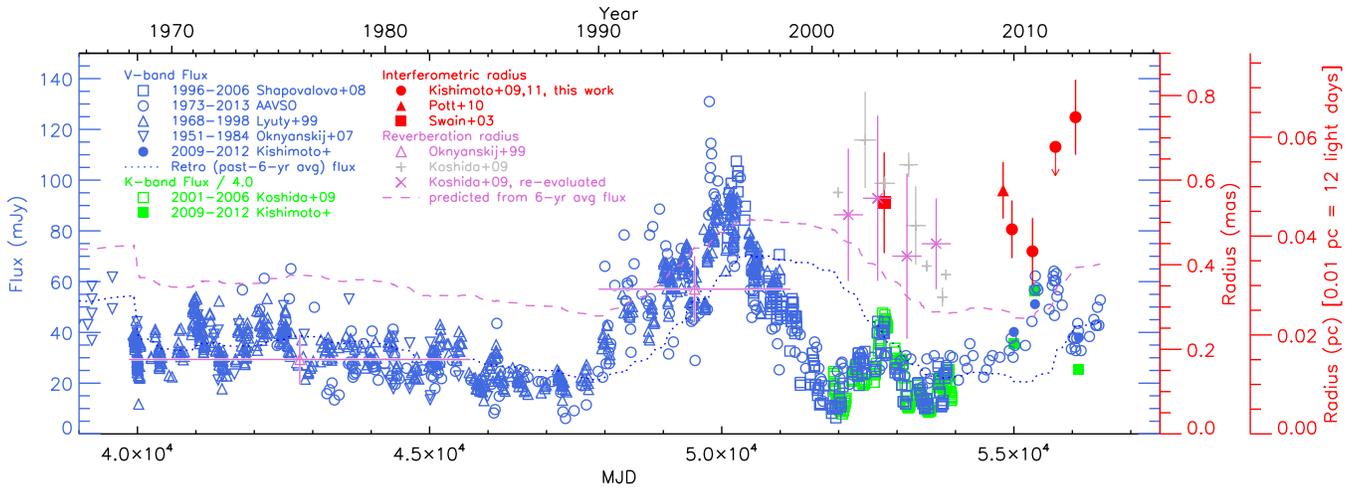}
\caption{Various measurements of the nuclear flux (scale on left axis) and the inner radius (scales on right axis) for NGC4151 as a function of time.
}
\label{fig_hysteresis_ngc4151}
\end{figure*}


\section{Results}

All the targets are partially resolved, except for NGC7469 where the observation was done when the system visibility seemed to be slightly unstable (bottom-left in Fig.\ref{fig_rawvis}). 
We fitted a thin-ring model to the observed visibility 
at each epoch in order to determine the implied overall size. This was done both before and after the small correction for the flux contribution from the accretion disk (AD) component, which is assumed to be unresolved. Here the AD flux fraction was estimated by fitting the SED with the AD and hot dust components (see Sect.3.1 in \citealt{Kishimoto11}). The fitted ring radii and AD fraction are summarized in Table~\ref{tab_obs}.

For the brightest Type 1 AGN in NGC4151, we have now accumulated $K$-band interferometric measurements at four epochs. Figure~\ref{fig_vis_ngc4151} directly compares the observed visibilities in the four epochs, together with two interferometric measurements in the literature \citep{Swain03,Pott10}. Here we also include our VLTI/AMBER measurements \citep{Kishimoto13}. 

Figure~\ref{fig_hysteresis_ngc4151} shows the fitted ring radii (after the AD correction) as a function of time. For 2011 and 2012 data, we included our VLTI data for the fit (Table~\ref{tab_obs}). In Fig.\ref{fig_hysteresis_ngc4151}, we also compiled several different photometric measurements: (1) continuum at 5117\AA\ from 1996 to 2006 by \cite{Shapovalova08} with their estimated host galaxy flux of 5.2 mJy subtracted; (2) AAVSO data\footnotemark[2], averaged over every 10 measurements in the $V$-band from 1973 to 2013; (3) $V$-band continuum from 1968 to 1998 by \cite{Lyuty99}; (4) photographic estimates in the $B$-band from 1951 to 1984 by \cite{Oknyanskij07}; (5) $V$-band continuum extrapolated from the AD+dust fit on our 0.9-2.2 $\mu$m SED in 2009/2010/2012 (\citealt{Kishimoto09KI,Kishimoto11}; Table~\ref{tab_obs}); (6) $K$-band continuum from 2001 to 2006 by \cite{Koshida09} and our $K$-band data taken in 2009/2010/2012 (both after the subtraction of the AD component), both scaled by a factor of 0.25, which is calculated from the average ratio of the $V$-band data of \cite{Shapovalova08} and $K$-band data of \cite{Koshida09}. 

We calculated a constant offset component in the optical coming from the host galaxy in the AAVSO data and the data from \cite{Lyuty99} and \cite{Oknyanskij07} using the data of \cite{Shapovalova08} as a reference and subtracted the averaged difference over the overlapping periods. We assumed a spectral shape of $f_{\nu} \propto \nu^{-0.3}$ for small corrections of non-$V$-band fluxes (\citealt{Francis91}; Fig.2 of \citealt{Kishimoto08}).

\footnotetext[2]{http://www.aavso.org}

First, we see that the observed visibilities and the corresponding implied ring-radii were rather stable from 2003 to 2010. This is despite the significant change in the central engine's flux, including a long-term increase from $\sim$2006, especially by a factor of a few from 2008 to 2010 \citep{Pott10,Kishimoto11}. Then, our interferometric observations in 2012 show evidence for a decrease of the visibility,
implying an increase in the radius 
of the dust sublimation region in NGC4151. The measured radius increase is at a $\sim$2.4 $\sigma$ level, which is the formal significance of the difference in the deduced radii calculated from the 1-$\sigma$ uncertainty of each radius determination (Table~\ref{tab_obs}). This increase in radius seems to have happened with an apparent delay of at least a few years from the significant flux increase seen especially from $\sim$2008 to 2011. 

For the other multiply observed targets, namely 3C273, MRK231, AKN120, and MRK509, we have not detected any significant variation in the size estimates in one to three year intervals. Finally, the sizes as a function of luminosity for the sample combined with other targets observed with the KI and also with the VLTI are discussed in \cite{Kishimoto13}.
Here we focus on the possible variation in visibility seen in NGC4151.

\section{Discussions}\label{sec_disc}

\subsection{Correlation with long-term average flux}

According to the accumulated interferometric observations of NGC4151 and the compiled flux measurements, the radius of the innermost dust distribution has apparently receded in response to the heating flux increase with a significant time delay. The observed delay implies that the overall timescale for the destruction of the innermost dust distribution is at least a few years.

In Fig.\ref{fig_hysteresis_ngc4151}, we also plotted near-IR reverberation radii.
These radii have overall increased from the 1970s (time lag of $\sim$18 days) to the 1990s ($\sim$35 days; \citealt{Oknyanskij99}) or the 2001-2006 period ($\sim$50 days; \citealt{Minezaki04,Koshida09}). The radius also seems to have decreased within the latter 2001--2006 period, which was found by \cite{Koshida09}. 

In \cite{Hoenig11}, we re-analyzed the data of \cite{Koshida09} and used the peak in the cross-correlation function to measure lags, rather than the centroid (which is adopted by Koshida et al.). This ensures the consistency with the other reverberation measurements by \cite{Oknyanskij99} and those implemented and compiled by \cite{Suganuma06}. The results are plotted in Fig.\ref{fig_hysteresis_ngc4151}. We found slightly shorter lags with larger uncertainties (see Sect.4 of \citealt{Hoenig11}), but the tendency of the decreasing radius is still roughly consistent with that found by \cite{Koshida09}.

The long-timescale increase of the radius from the 1970s to the 1990s and its subsequent decrease possibly seen in the 2001-2006 period could well be due to the historical flux increase peaking around 1996 (Fig.\ref{fig_hysteresis_ngc4151}) and the subsequent significant decrease in the flux. The former probably destroyed grains in the innermost region, and the latter lead to the reformation of the dust grains in the same region. If this is the case, there must have been quite a long delay of several years between the decrease in flux and that in the reverberation radii measured in the 2001-2006 period. This could, in principle, constrain the dust reformation timescale, implying that it might have taken several years for the innermost dust grains within a $\sim$0.05~pc radius region to re-form. This reformation timescale was indeed already inferred by \cite{Oknyanskij06}.\footnotemark[3]

\footnotetext[3]{
Their inference was based on an additional, interesting measurement of the near-IR time delay of 104$\pm$10 days for the period 1999-2004 shown in the conference paper \citep{Oknyanskij06}. This would further support our conclusions, but with the quoted uncertainty, the measured lag seems slightly inconsistent with those from the data of \cite{Minezaki04} and \cite{Koshida09} for the overlapping period (see Fig.\ref{fig_hysteresis_ngc4151}). A further evaluation of the uncertainty would be desirable.}

\begin{figure*}
\includegraphics[width=9.0cm]{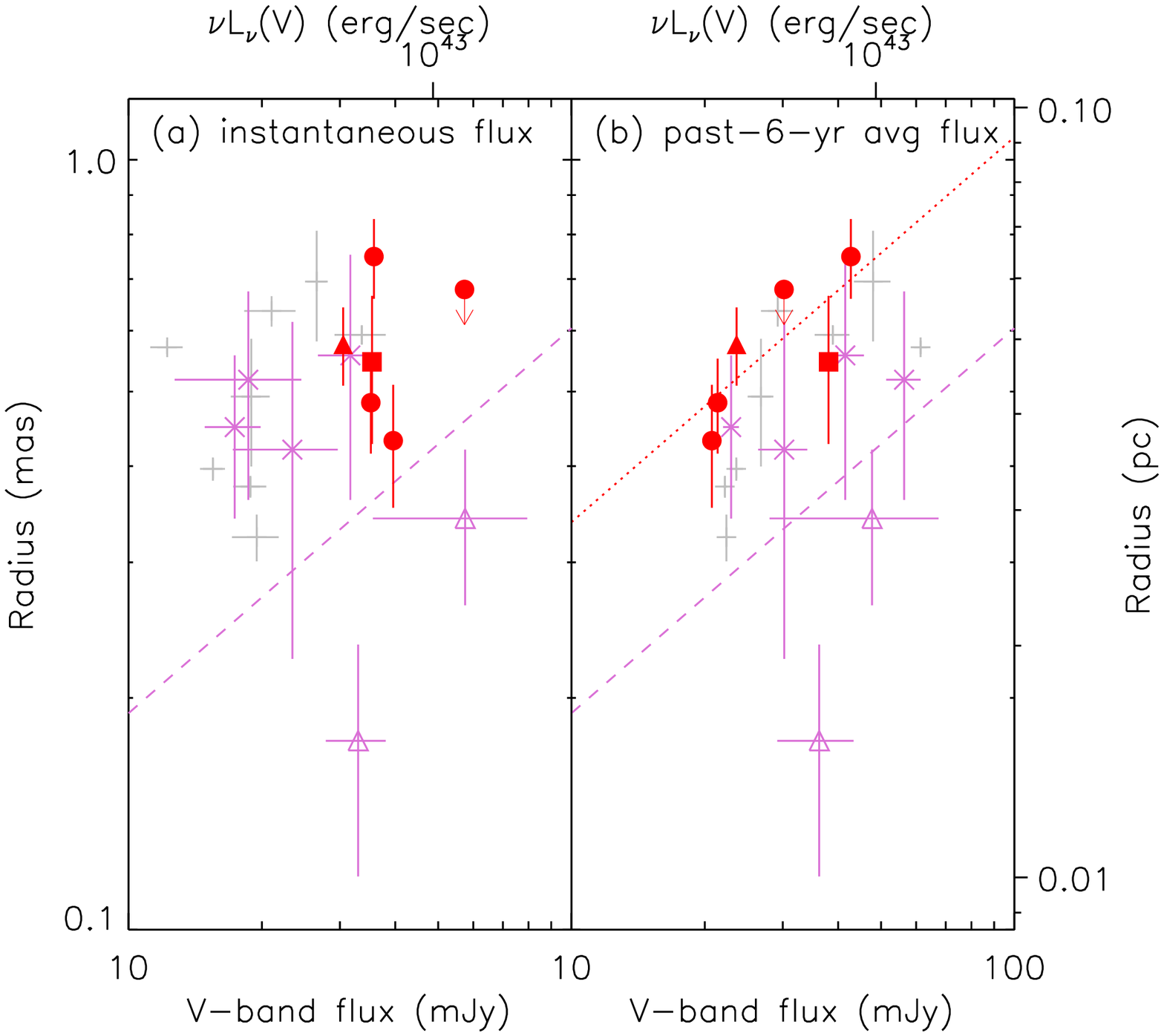}%
\hspace{0.3cm}%
\includegraphics[width=8.7cm]{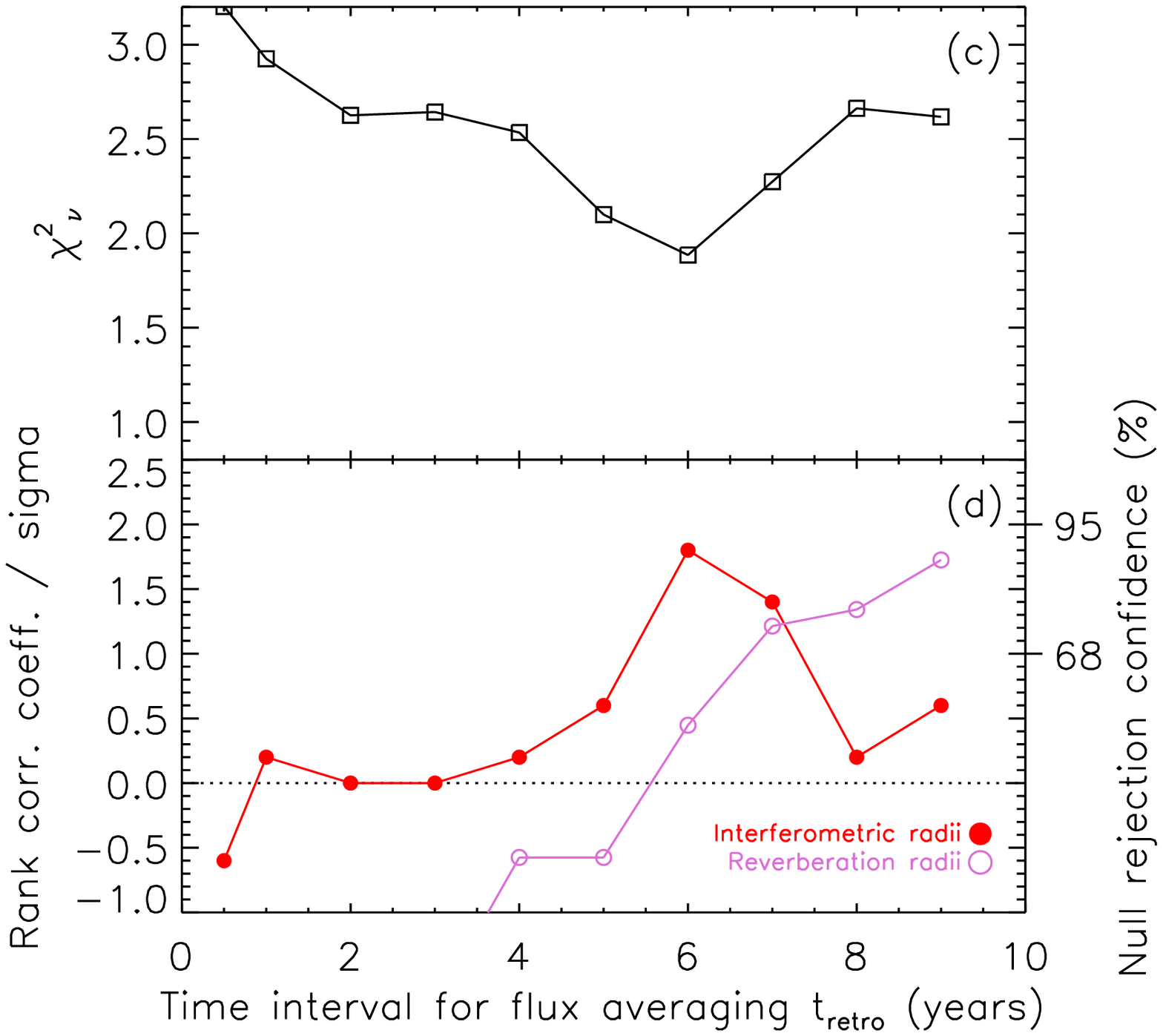}
\caption{Various radius measurements versus ($a$) instantaneous flux and ($b$)  
'retro' flux with $t_{\rm retro}$ of six years, which is the average flux over the past six years from each radius measurement. Symbols are the same as in Fig.\ref{fig_hysteresis_ngc4151}. 
The dashed line is the $L^{1/2}$ fit by \cite{Suganuma06} for the $K$-band reverberation radii.
In the right panels, statistics are shown as a function of $t_{\rm retro}$ in
($c$) reduced $\chi^2_{\nu}$ for the deviation of the radius measurements from the $L^{1/2}$ proportionality (normalization for the interferometric radii is separately determined by a fit, shown in the dotted line in panel $b$),
and ($d$) rank correlation coefficients in units of the standard deviation expected in the null hypothesis.
}
\label{fig_radlum_ngc4151}
\end{figure*}

Note that here we talk about the destruction of the dust {\it distribution}, 
but not necessarily 
the microscopic destruction of each grain. This distribution radius does not seem to be changing sensitively to the {\it instantaneous} change in the incident heating radiation (\citealt{Pott10,Kishimoto11,Hoenig11}). This is also shown in Fig.\ref{fig_radlum_ngc4151}$a$, where we plot the measured radii against the instantaneous incident flux (in practice, we calculated the flux averaged over the past half a year from the radius measurement). The radius does not show the expected $L^{1/2}$ dependence (as already found by \citealt{Koshida09}), which is given by the relationship between the reverberation radius and luminosity over a {\it sample} of objects \citep{Suganuma06} and shown as the dashed line in Fig.\ref{fig_radlum_ngc4151}$a$. This probably means that the innermost dust grains are in the optically-thick accumulations or clouds so that only the dust grains at the surface region sublimate first \citep{Hoenig11}. This would slowly be followed by the increase of the overall radius of the distribution when the incident flux level is persistently kept high.

One simple model to quantify the possible destruction and reformation of the dust distribution is to postulate that the distribution radius is approximately determined by the average of the incident flux over some past period, or an average 'retro' flux. The retro flux at a given observing time of a radius measurement is set by averaging the optical flux over the past $t_{\rm retro}$ years back from the radius observing time. The reformation time and destruction time of the dust distribution are not necessarily the same, with the former likely being longer than the latter. Here, we take the simplest first approximation to use a single averaging timescale $t_{\rm retro}$, which can be regarded as an overall mean of the two timescales. 

Calculating the deviation from the square-root proportionality and also the Spearman's rank correlation coefficients between the radius and retro flux, we searched for the best value of $t_{\rm retro}$ for the compiled flux and radius measurements, finding it to be $\sim$ 6 years (Fig.\ref{fig_radlum_ngc4151}$c,d$). Here we calculated the deviation from the square-root proportionality for reverberation radii using the $L^{1/2}$ fit by \cite{Suganuma06}, and the deviation for interferometric radii by fitting a $L^{1/2}$ relation separately. This is because we anticipate the latter radii to be slightly but systematically larger than the former. More specifically, the reverberation measurement using the peak of the cross-correlation function tends to probe the fastest responding material \citep{Koratkar91}, i.e. give a radius close to the inner boundary radius, while the interferometric radius represents an overall average radius of the brightness distribution, thus slightly larger than the reverberation radius. This difference between the interferometric and reverberation radius has been seen in other different objects \citep{Kishimoto11}. Similarly, we calculated the rank correlation coefficient separately for the two different radii. 

Figure~\ref{fig_radlum_ngc4151}$b$ shows the radii as a function of the retro flux for the case of $t_{\rm retro}$ = 6 years. The reduced $\chi^2_{\nu}$ for the deviation from the square-root proportionality has a minimum of 1.8 at this $t_{\rm retro}$ (Fig.\ref{fig_radlum_ngc4151}$c$). The confidence level of rejecting a null hypothesis for the correlation between the interferometric radii and the retro flux has a peak of 93\% (or 1.8 $\sigma$) at this $t_{\rm retro}$, while the rank correlation coefficient for the reverberation radii becomes positive only at $t_{\rm retro} \gtrsim 6$ years though not with a high confidence level  (Fig.\ref{fig_radlum_ngc4151}$d$). All these indicate that the radius is relatively well correlated with the {\it long-term average} of the incident flux rather than the instantaneous flux, and consistent with being proportional to the square-root of the {\it long-term average} luminosity. 
Finally, Figure~\ref{fig_radlum_ngc4151}$b$ also suggests that the interferometric radius is larger than the reverberation radius by a factor of $\sim$1.8 for a given retro luminosity in this AGN.


\cite{Koshida09} have already pointed out that, in their light curves between 2001 and 2006, the reverberation radius in the early part ($\sim$2001) seems to be affected by the past bright flux. Indeed, we see the historical peak around 1996 (see Fig.\ref{fig_hysteresis_ngc4151}). Therefore, it is better to put the data into a perspective covering a much wider time range. With the analysis done here, we indeed find a likely timescale of $\sim$6 years, rather than the one-year time scale suggested by \cite{Koshida09}. This is depicted in Fig.\ref{fig_radlum_ngc4151}$c$ and $d$.

\subsection{Physical implications}

We can relate this dust timescale $t_{\rm retro}$ to the radial UV optical thickness of the sublimation region, or more specifically, $\tau_{\rm UV}$ over the inner radial length scale of $\sim$0.05~pc in this AGN. This is coarsely given as $\tau_{\rm UV} \simeq t_{\rm retro} / t_{\rm dest}$, where we denote the (microscopic) dust destruction timescale as $t_{\rm dest}$, within which the region of UV optical thickness unity will be destroyed when flux is increased. The dust evaporation occurs when the dust vapor pressure dominates the ambient dust gas pressure, and the destruction timescale $t_{\rm dest}$ can roughly be estimated as a function of the dust vapor pressure $P_{\rm vap}$ and the dust temperature $T$. This is given as (see e.g. \citealt{Phinney89,Guhathakurta89,Kimura02,Kobayashi11})
\begin{eqnarray}
t_{\rm dest} & \simeq &
\frac{4}{3} \pi a^3 \rho
\left[ 4 \pi a^2 \sqrt{ \frac{m}{2 \pi k T} } P_{\rm vap} \right]^{-1}\\
& \simeq & 4.7 \times 10^{1}
\left( \frac{a}{0.1 \ \mu {\rm m}} \right)
\left( \frac{T}{T_0} \right)^{1/2}
\left( \frac{P_{\rm vap}}{P_0} \right)^{-1}
\ {\rm (days)}, 
\label{eq_tdest}
\end{eqnarray}
where $a$ is the radius of dust grains assumed to be spherical, $m$ the dust molecular mass, $k$ the Boltzmann constant, and $\rho$ the bulk density of a dust particle. 

Here we normalized $T$ by $T_0$=1400~K, which is the typically observed color temperature of the dust SED in the near-IR and adopted here as a proxy for the actual temperature of sublimating grains (see more below on the temperature). We used the values of $m$ and $\rho$ for silicate grains, but adopting the $m$ and $\rho$ values for other grains does not change the order of magnitude of the estimated $t_{\rm dest}$. 

The estimation is more sensitive to vapor pressure $P_{\rm vap}(T)$. We normalized this by the constant $P_0$$\equiv$$n_0 k T_0$, with $n_0$=$10^6$ cm$^{-3}$, i.e. $P_0 \equiv 2\times 10^{-7}$ dyn cm$^{-2}$ (see more below on the density), because silicate grains give $P_{\rm vap}(1400{\rm K})$$\sim$$P_0$, adopting constants from the literature \citep{Guhathakurta89,Kimura02}. The vapor pressure as a function of $T$ is given as $P_{\rm vap} \propto e^{-A/T}$, where the constant $A$ depends on the dust species and $A$$\sim$$6.8 \times 10^4$~K for silicate grains, and is thus, very sensitive to $T$. However, we can still use a fixed temperature for this estimation as long as $t_{\rm dest}$ is shorter than the flux variation timescale, which seems to be the case here (see Eq.\ref{eq_tdest}). Note that the observed near-IR color temperature is probably a lower limit for the actual temperature of sublimating grains, and thus $t_{\rm dest}$ can even be shorter.

Thus, we have
\begin{equation}
\tau_{\rm UV} \simeq 4.7 \times 10^{1}
\left( \frac{t_{\rm retro}}{\rm 6 \ years} \right)
\left( \frac{a}{0.1 \ \mu {\rm m}} \right)^{-1}
\left( \frac{T}{T_0} \right)^{-1/2}
\left( \frac{P_{\rm vap}}{P_0} \right).
\end{equation}
Therefore, the region is estimated to be very optically thick over the radial length scale of $\sim$0.05~pc. This is consistent with the general expectation from the optically thick equatorial obscuration. Conversely, in order to have such a high optical thickness, the destruction timescale $t_{\rm dest}$ must be tens of days (recall $\tau_{\rm UV} \simeq t_{\rm retro} / t_{\rm dest}$), and this requires the vapor pressure $P_{\rm vap}$ to be on the order of $P_0$. This corresponds to the sublimation at $\sim$1400~K in the case of silicate grains. In the case of graphite grains, adopting constants again from the literature \citep{Phinney89,Guhathakurta89}, they should be sublimating at a higher temperature of $\sim$1800~K, at which they have $P_{\rm vap}$$\sim$$P_0$.

This vapor pressure in the sublimation region also implies that the ambient density of gas-phase dust constituents is on the order of $\sim$10$^6$ cm$^{-3}$ (denoted as $n_0$ above), corresponding to the hydrogen number density of $\sim$10$^9$ cm$^{-3}$ for the normal ISM abundance. It is worth noting that this density is comparable to that estimated for the broad-line region.

%

\subsection{Concluding remarks}

It is certainly true that the inference and discussions here are based on a limited number of radius measurements. Further near-IR interferometric observations of this well-monitored object will certainly be yielding. Based on the available optical light curve, we can actually predict that such measurements in the coming few years will show a large radius similar to what we measured in 2012. These and further measurements will better constrain the dust-related timescales and will lead to new physical insights and advance our understanding of the physics and structure of the innermost dusty region.




\acknowledgements



The data presented herein were obtained at the W.M. Keck Observatory, which is operated as a scientific partnership among the California Institute of Technology, the University of California and the National Aeronautics and Space Administration (NASA). 
This work has made use of services produced by the NASA Exoplanet Science Institute at the California Institute of Technology.  
We thank the Calar Alto Observatory for allocating director's discretionary time to this program. We acknowledge with thanks the variable star observations from the AAVSO International Database contributed by observers worldwide and used in this research.





\end{document}